# The Nosoi commute: a spatial perspective on the rise of BSL-4 laboratories in cities

**Authors:**

Thomas P. Van Boeckel[1,2*], Michael J. Tildesley[3], Catherine Linard[2,4], José Halloy[5], Matt J. Keeling[3] and Marius Gilbert[2,4]

**Affiliations:**

1. Ecology and Evolutionary Biology Department, Princeton University, Princeton, NJ, USA.

2. Biological Control and Spatial Ecology Lab, Université Libre de Bruxelles, Brussels, Belgium.

3. Mathematics Institute, University of Warwick, Coventry, UK.

4. Fonds National de la Recherche Scientifique, Brussels, Belgium.

5. Paris Interdisciplinary Energy Research Institute, University Paris Diderot, Paris, France.

*Correspondence to: thomas.van.boeckel@gmail.com

**Abstract**

Recent H5N1 influenza research has revived the debate on the storage and manipulation of potentially harmful pathogens. In the last two decades, new high biosafety (BSL-4) laboratories entered into operation, raising strong concerns from the public. The probability of an accidental release of a pathogen from a BSL-4 laboratory is extremely low, but the corresponding risk -- defined as the probability of occurrence multiplied by its impact -- could be significant depending on the pathogen specificities and the population potentially affected. A list of BSL-4 laboratories throughout the world, with their location and date of first activity, was established from publicly available sources. This database was used to estimate the total population living within a daily commuting distance of BSL-4 laboratories, and to quantify how this figure changed over time. We show that from 1990 to present, the population living within the commuting belt of BSL-4 laboratories increased by a factor of 4 to reach up to 1.8% of the world population, owing to an increase in the number of facilities and their installation in cities. Europe is currently hosting the largest population living in the direct vicinity of BSL-4 laboratories, while the recent building of new facilities in Asia suggests that an important increase of the population living close to BSL-4 laboratories will be observed in the next decades. We discuss the potential implications in term of global risk, and call for better pathogen-specific quantitative assessment of the risk of outbreaks resulting from the accidental release of potentially pandemic pathogens.

## Introduction

In recent years there has been a huge proliferation in the study of pathogens, which has promised many breakthroughs in human health. This has led to several new high biosafety (designated Biosafety Level 4 or BSL-4) laboratories entering into operation [1,2]. However, experimentation with pathogens is not without risk. There have been strong concerns from the general public [1] and the scientific community [3–6] over the handling of potentially deadly human pathogens, in part fuelled by the recent work on H5N1 influenza [7,8]. A recent study quantified the risk that an accidental release of such pathogen could not be contained in the local community, and showed that this would be strongly influenced by the vicinity of the laboratory in terms of human population, i.e. that the risk would be higher in urban than rural context [9]

The probability of the release of a pathogen from one of the highest biosafety laboratories can be considered to be extremely low [10] and is in theory comparable for all BSL-4 laboratories. All facilities follow standardized criteria and use similar materials and resources to enable them to operate at the highest security level. However, this is nothing exceptional [11]. An interesting precedent in risk assessment of potentially dangerous scientific research was set by an experiment carried out at the Large Hadron Collider (LHC). The probability that the experiment could create black holes during its operation was seriously evaluated, because of its potentially devastating consequences, despite the belief that the probability of such an event occurring was extremely low [12]. Similarly, the assessment of security in nuclear power plants also involves extremely low probabilities of events, but is evaluated extremely carefully; the recent example of Fukushima highlights the dramatic consequences of an unexpected sequence of contingencies. Leaks in high biosafety laboratories have occurred in the past [11,13,14], some of which have resulted in local contagion [14] and could have resulted in large-scale epidemics. In a first effort to better characterize this risk, we quantified how the population living in the vicinity of BSL-4 laboratories has changed over time.

A list of existing BSL-4 facilities was assembled from publicly available sources of information including the United Nations Biological Weapons Convention (Data and Methods section). The list included the geographical coordinates (Fig 1, A) of each facility and the date it started its operations. The next step involved evaluating the size of the population that lives in the vicinity of each laboratory representing a potential biological hazard. We considered the hypothetical situation where a lab-worker is accidentally infected to estimate the population living within the commuting belt for this worker. Specifically, we estimated the population size living within a typical 30-minute commute (15, 16) of each laboratory (Fig 1 B-D). The global population living in the direct vicinity of BSL-4 laboratories was then defined as the total population living within the commuting belts of all facilities.

## Data and Methods

### *List of BSL-4 Laboratories*

A list of existing BSL-4 facilities was assembled from publicly available sources of information such as institutions and non-governmental organization (NGO) websites, scientific publications [15] national newspapers and the archives of the United Nations Biological Weapons Convention. The list included the name, coordinates and period at which the facility entered in operation. For simplicity and because there was some uncertainty in some of the dates, four periods were identified, before 1990, 1990 to 2000, 2000 to 2010 and after 2010. When an opening year could not clearly be identified, different sources were crossed to identify the period during which the facility opened (See column Operational Date in Supplementary Table 1). Out of 55 listed laboratories, three (Veterinary Laboratories Agency, United Kingdom, Republican Research and Practical Center for Epidemiology and Microbiology Belarus and Preventive Medical Institute of the Ministry of National Defence, Taiwan) could not be assigned a starting period because of insufficient information. These three laboratories were therefore excluded from the analysis; although they may still exist.

The list established is undoubtedly incomplete with regards to all facilities suspected to exist, because countries do not all communicate with an equivalent level of transparency regarding their research activity on dangerous pathogens. However, in the absence of an official and transparent list of BSL-4 facilities maintained at the international level, the present list may be considered as the most up-to-date source of information. Finally, several countries distinguish between facilities operating on human or animal pathogens. However, recent research on influenza indicates that this discrimination is obsolete for a range of pathogens, and therefore the BSL-4 laboratories described in this study include both types of facilities.

The authors stand ready to update the list established with any information arising from the concerned institutions regarding localization or dates when facilities entered in operation, and to re-evaluate their estimates accordingly.

*Commuting Belts and Demography Maps*

A friction surface was used to delineate a commuting belt of 30 minutes around each laboratory. In this case, the friction surface used contained the value in minutes required to cross a 1 kilometer pixel [16]. The time to cross each pixel from a friction surfaces is calculated from maps of environmental and anthropogenic variables such as de type of land use, transport networks elevation, slope etc. Using a cumulative sum function combined with such a surface allows us to calculate an isochronal belt reachable for a hypothetical lab worker commuting home on a 30 minutes journey. These commuting belts were then used to sum up the population in the direct vicinity of each laboratory, as reported on figure 2 and 3. A sensitivity analysis using commuting duration of 10 to 60 minute was conducted to insure the consistency of the pattern observed across a range of plausible commuting values (SI Fig. 1,2).

The threshold value of 30 minutes was chosen since it lies within the observed range of values for developed countries across the different periods: according to different sources the average commuting time in the US in 2009 was 25.1 minutes [17] and 37.5 minutes in western Europe in 2000 [18].

The demography maps used were obtained from the *Global Rural Urban Mapping Project* [19] population database for the years 1990, 2000 and 2010. The demography estimates for the year 2010 were used both for 2010 and the post 2010 period as most laboratories expected to be built after 2010 and included in this study are due in 2012.

All the analyses were carried out in the statistical programming language R (***cran**.**r**-project.org/)* and the maps produced with *ArcGIS 9.3 (www.esri.com).*

## Results

Our findings showed that the global population living within 30 minutes of BSL-4 laboratories increased from 30,165,678 in 1990 to 42,456,931 in 2000 and to 96,986,631 in 2010. Prediction based on facilities built since 2010 or currently under construction suggested that this figure should increase up to 126,146,118 after 2012. Overall, this represented a 4-fold increase from 0.57% of the world population in 1990 to 1.8% after 2012.

The dramatic increase in the total population living in the immediate vicinity of BSL-4 laboratories was primarily due to the increase in the number of laboratories (12 in 1990, 17 in 2000, 42 in 2010 and 52 after 2012). Comparatively the population growth around the existing laboratories, only accounted for 5.2% of the increase since 1990 (Fig 2). The construction of new facilities in densely populated areas played a key role in the predicted rise in the population exposed. A sensitivity analysis on the commuting time between 10 and 60 minutes showed these figures to range from 29,040,972 to 246,272,658 people for the post 2012 period. Interestingly, we find that smaller commuting belts (10 minutes) contain more individuals than would be expected from a simple geometric argument (eg a 30-minute commuting belt contains less than 9 times the number of individuals within the 10-minute belt). The ratios of population between the different commuting belts were respectively: $P_{30min}/P_{10min} =$ 4.29 instead of 9 and $P_{60min}/P_{30min}$ = 1.97 instead of 4. This suggests that laboratories tend to be located in the locally highest population densities. This trend is also illustrated by Figure 3a -- whilst there are far fewer BSL-4 laboratories in Asia than in North America, there are a larger number of people living in the immediate vicinity of these laboratories. Europe hosts the largest number of laboratories and because of its densely populated landscape; it also has the largest population of people living in the commuting belts of these facilities. Figure 3b shows how the top 10 facilities having the largest population in their commuting belts have changed over the last two decades. The situation in 1990 reflected the historical context at the end of the cold war, with five laboratories in the top 10 located in NATO countries and a further three in the USSR. By 2000, nine out of ten laboratories with the largest population in their commuting belt were in the western world, with the 10$^{th}$ lab being located in South Africa. By 2010, new facilities had been constructed in densely-populated areas in Europe (London, Milan, Hamburg) and in Asia (Taiwan, Singapore). According to the predictions for the post 2010 era, India will make a noticeable entry in this ranked list, with the country's first two BSL-4 facilities being built in Pune (5.5 million inhabitants) and Bhopal (1.8 million inhabitants). Meanwhile North America only had one facility left in this top 10 in 2010: the NIH in Bethesda, Maryland, USA. Interestingly, in all four periods the United Kingdom was the nation with by far the highest population living in the vicinity of BSL-4 laboratories. This stemmed both from the record number of BSL-4 facilities in the country (9, in 7 sites) and their distribution in and around the capital city of London, the largest city in Europe.

## Discussions

Even assuming a constant very low probability per laboratory, the global risk of an accident has increased owing to the proliferation of BSL-4 laboratories. In addition, new facilities were mostly established in high-density urban areas (Fig. 1A), although the impact of this on the combined risk is more difficult to assess. However, recent results of simulation models suggest that urbanization of BSL-4 laboratories would indeed increase the risk that an accidental release could not be contained [9]. The total population of people living in the vicinity of BSL-4 laboratories is one of several factors that may affect the chance that an accidentally released pathogen would trigger an epidemic. A comprehensive quantification of this risk would require a robust and complex pathogen-specific epidemic model accounting for epidemiology, age structure, contact rates, transport networks, intervention and diagnosis capacities of each country hosting a BSL-4 laboratory [20]. Since, experiments on potentially pandemic pathogens such as influenza or SARS are currently also authorized in BSL-3/3+ laboratories, such pathogen-specific assessment should also include the BSL-3/3+ facilities that have engaged on research on those pathogens. Instead we have adopted a simple approach, focusing on BSL-4 laboratories, and quantifying the local population in their immediate vicinity. This resonates with the intuitive understanding that the consequences of an infectious disease agent may be very different should it escape a laboratory located in cities like London or Bhopal as opposed to remote areas such as Těchonin in the Czech Republic or in the Rocky Mountains in the USA.

Research on potentially dangerous disease agents has many scientific and societal benefits; however these must always be weighed against their low-probability but high-impact risks. The recent multiplication of BSL-4 laboratories, not to mention BSL-3 laboratories that are far more numerous and harder to identify, raises one key question. Can the multiplication of the number of laboratories and their installation in densely populated areas cause a substantial increase in the risk of a man-triggered epidemic or pandemic? The results presented in this paper indicate that this may indeed be the case. Whilst competition between research groups and countries is a stimulating factor in research, there is the possibility for unnecessary repetition of potentially dangerous experiments and hence an associated replication of risk. The current situation, whereby new BSL-4 facilities tend to be located in regions of high population density, may ultimately result in the risks of an artificial outbreak occurring outweighing the risk of a naturally-arising global pandemic, as recently discussed in several opinion papers [21,22]. The scientific community and policy makers therefore need to strike a careful balance between scientific competition, national independence and global risk. Better international cooperation and harmonization of regulation in this very particular field of research could have both an immediate and substantial impact on the risk of future outbreaks.

**Funding Source:**

This work was supported by the Belgian *Fonds National pour la Recherche Scientifique*

**Acknowledgments:**

The authors are grateful to *Leon Danon* (Warwick Mathematics Institute*),* for stimulating discussions on the topic and to *Aiko Gryspeirt* (Université Libre de Bruxelles) for her help with the BSL-4 data localization.

**Figure 1**

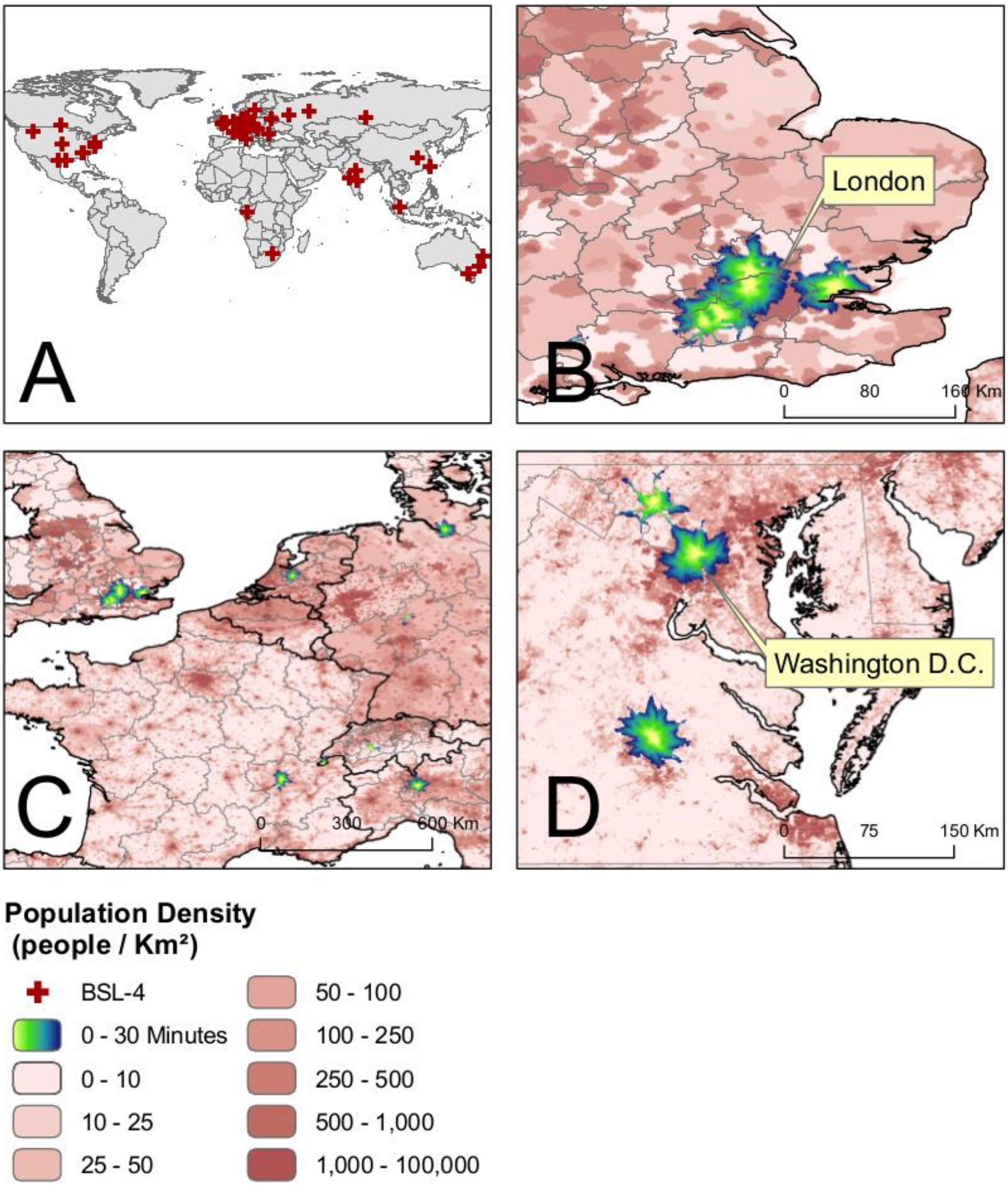

**Distribution of BSL-4 Laboratories and population**. Global distribution of Biosafety Level 4 Laboratories (A). Population density and commuting belts of 30 minutes around Biosafety Level 4 Laboratories in Western Europe (B) South of England (C) and East Cost of the United States (D).

**Figure 2**

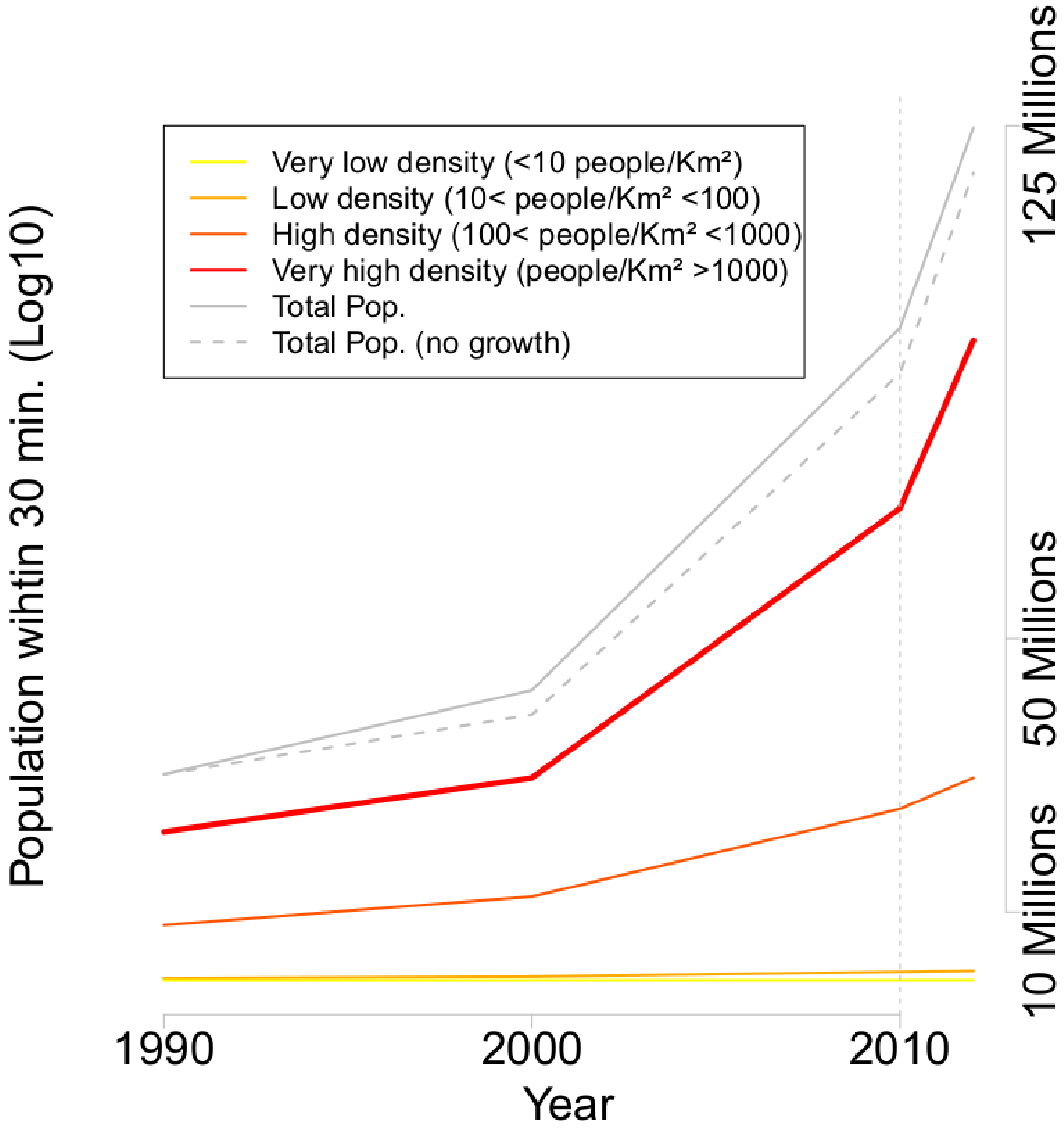

**Global population living in the immediate vicinity of BSL-4 laboratories since 1990.** The yellow to red lines highlights that a significant part of this increase is due to the fact that new laboratories were established in densely populated urban areas. The dashed grey line shows that subtracting the growth in human population during the last two decades has a negligible effect on the increase of population.

**Figure 3**

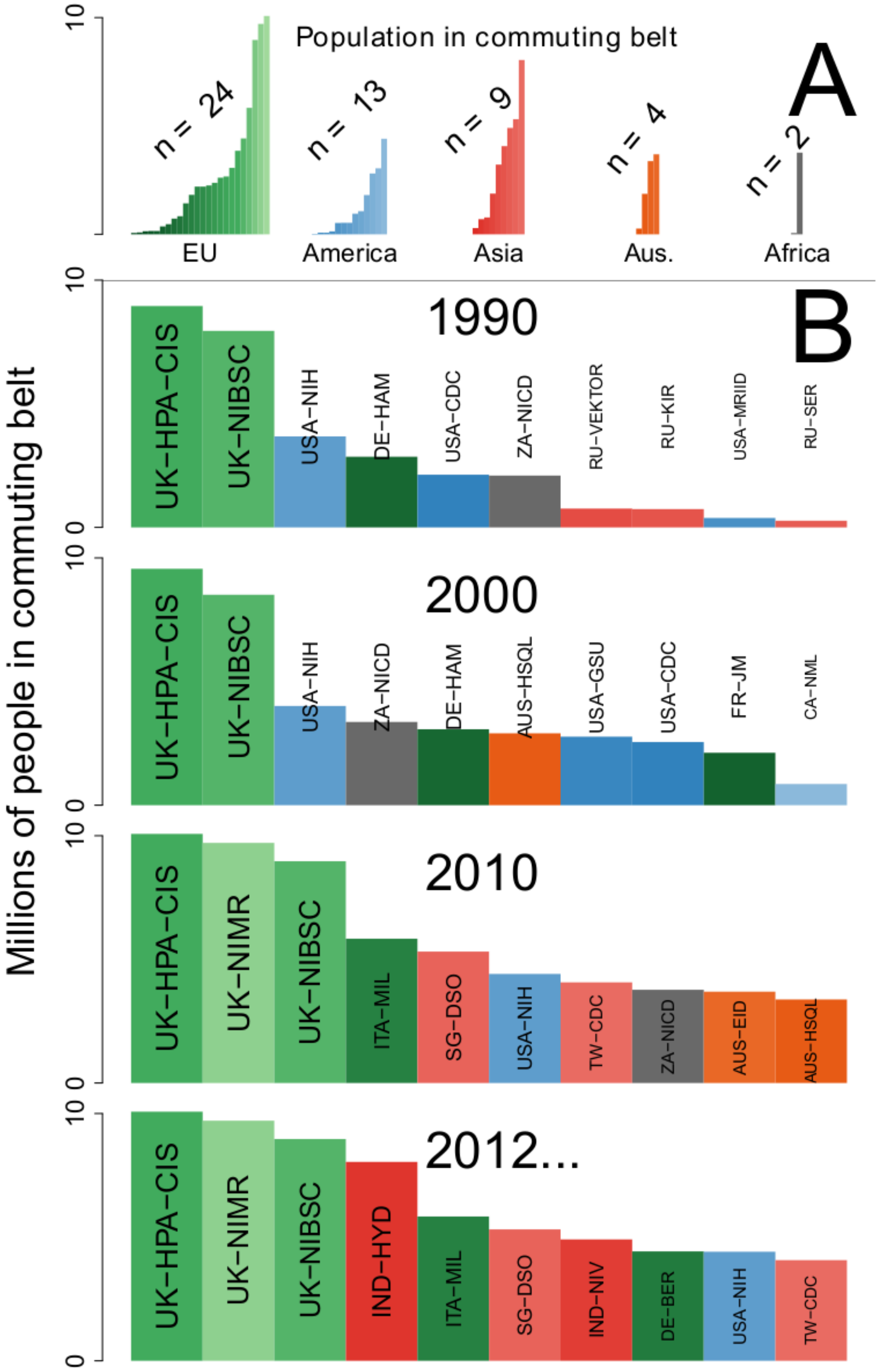

**Regional trends in the evolution of the population living in the immediate vicinity of BSL-4 laboratories since 1990.** (A) Distribution of population within BSL-4 laboratories commuting belts by region after 2012. (B) Evolution of the top 10 BSL-4 hosting the largest population in their commuting belt since 1990. *(USA United States; UK United Kingdom; ZA, South Africa; RU Russia (and previously USSR); SG Singapore; TW Taiwan; AUS Australia; ITA Italy; IND India; DE Germany; FR France; CA Canada)*

Supplementary Figure 1

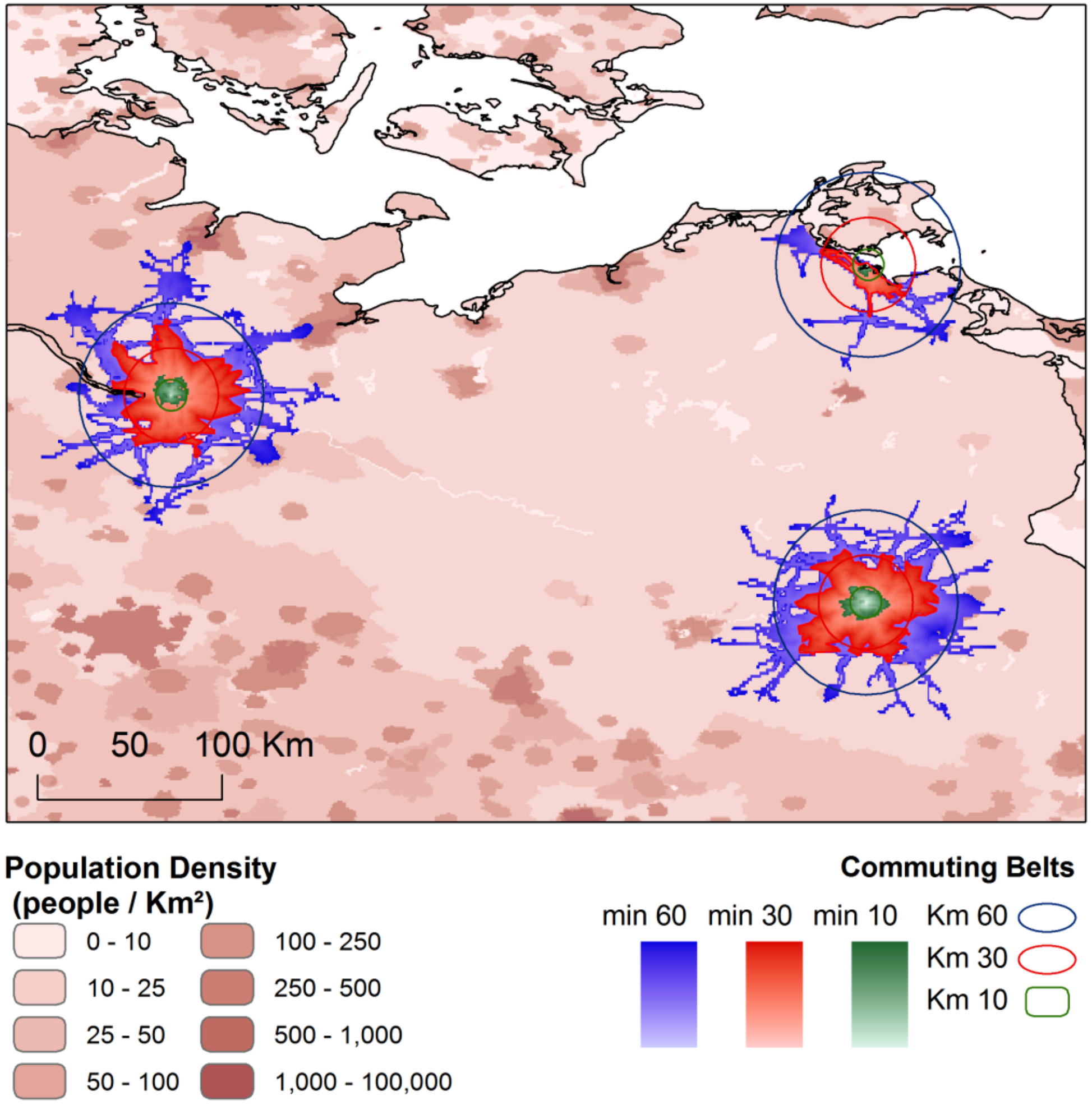

**Comparative map of commuting belts around Biosafety Level 4 Laboratories in Northern Germany.** Colors scale indicates the duration/distance of the commute for 10 (green), 30 (red) and 60 (blue) minutes/kilometers.

Supplementary Figure 2

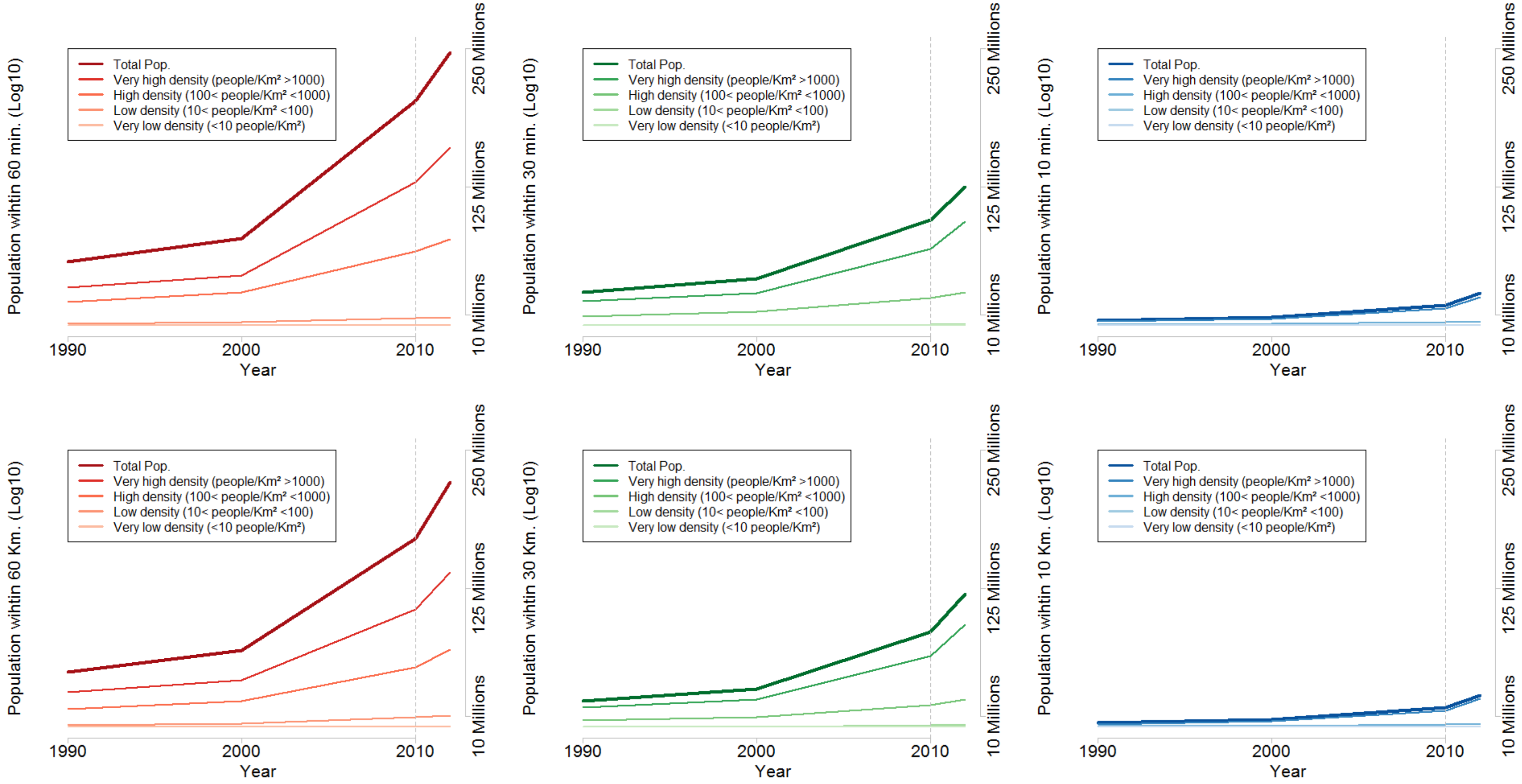

**Global population living in the immediate vicinity of BSL-4 since 1990.** Millions of people by population density classes for commuting belts of 60, 30, 10 minutes (top) and 60, 30, 10 kilometers (bottom)

## Supplementary Table 1. Current List of BSL-4 Facilities*

| Institution | Code | Location | Country | Operational Date |
|---|---|---|---|---|
| Centers for Disease Control and Prevention | USA-CDC | Georgia, Atlanta | USA | 1988 |
| Center for Biotechnology and Drug Design Georgia State University | USA-GSU | Georgia, Atlanta | USA | 1994 |
| Division of Consolidated Laboratory Services | USA-DCLS | Virginia, Richmond | USA | 2003 |
| United States Army Medical Research Institute for Infectious Diseases | USA-USAMRIID | Maryland, Fort Detrick | USA | 1969 |
| National Biodefense Analysis and Countermeasures Center (NBACC) | USA-NBACC | Maryland, Fort Detrick | USA | 2008 |
| Integrated Research Facility | USA-IRF | Maryland, Fort Detrick | USA | 2009 |
| National Institutes of Health (NIH) | USA-NIH | Maryland, Bethesda | USA | <1985 |
| National Bio and Agro-Defense Facility (NBAF) | USA-NBAF | Manhattan, Kansas | USA | 2020 |
| NIAID Rocky Mountain Laboratories | USA-NIAID | Montana, Hamilton | USA | 2008 |
| Galveston National Laboratory, National Biocontainment Facility | USA-GNL | Texas, Galveston | USA | 2008 |
| Center for Biodefense and Emerging Infectious Diseases Shope Laboratory | USA-SHOPE | Texas, Galveston | USA | 2003 |
| Texas Biomedical Research Institute (Southwest Foundation for Biomedical Research) | USA-TBRI | Texas, San Antonio | USA | 2000 |
| National Microbiology Laboratory | CA-NML | Manitoba, Winnipeg | Canada | 1999 |
| Australian Animal Health Laboratory (AAHL) | AUS-AAHL | Victoria, Geelong | Australia | 1985 |
| National High Security Laboratory (NHSQL); Victorian Infectious Disease Reference Laboratory | AUS-NHSQL | Victoria, North Melbourne | Australia | 1996 |
| Virology Laboratory of the Queensland Department of Health | AUS-VLQ | Queensland, Coopers Plains | Australia | >2000 |
| Emerging Infectious Diseases and Biohazard Response Unit | AUS-EIDBRU | Westmead | Australia | 2007 |
| Wuhan Institute of Virology of the Chinese Academy of Sciences | CN-WUHAN | Hubei, Wuhan | China | 2010 |
| Centre for Cellular and Molecular Biology | IND_HYD | Hyderabad | India | 2010 |
| National Institute of Virology, Indian council of medical research | IND-NIV | Pune | India | 2012 |
| High Security Animal Disease Laboratory (HSADL) | IND-BO | Bhopal | India | 2000 |
| State Research Center of Virology and Biotechnology VECTOR | RU-VEKTOR | Novosibirsk Oblast, Koltsovo | Russia | <1990 |
| Institute of Microbiology | RU-KIR | Kirov | Russia | <1990 |
| Virological Center of the Institute of Microbiology | RU-SER | Sergiev Possad | Russia | <1990 |
| Defence Science Organization | SG-DSO | Singapore | Singapore | 2003 |
| Kwen-yang Laboratory Center of Disease Control (Taiwan) | TW-CDC | Taipei, Taiwan | Taiwan | >2003 |
| Preventive Medical Institute of ROC Ministry of National Defense | TW-PMI | Taiwan | Taiwan | <2003 |
| Republican Research and Practical Center for Epidemiology and Microbiology | BL-RRPCEM | Minsk | Belarus | <2000 |
| Army Center for Medical Research | ROM-AMR | Romania | Romania | >2011 |
| Laboratory for Biological Monitoring and Protection | CZR-KAM | Kammena | Czech Rep. | 2007 |
| State Veterinary Institute Prague | CZR-PRA | Prague | Czech Rep. | 2007 |
| Biological Defense Center | CZR-TEC | Techonin | Czech Rep. | >2005 |
| National Center for Epidemiology | HUN-NCE | Hungary | Hungary | 2002 |
| Laboratoire P4 Jean Mérieux | FR-JM | Rhône-Alpes, Lyon | France | 1999 |
| Bernhard Nocht Institute for Tropical Medicine | DE-HAM | Hamburg | Germany | <1987 |
| Friedrich Loeffler Institute on the Isle of Riems | DE-FLI | the Isle of Riems (Greifswald) | Germany | 2011 |
| Philipps University of Marburg | DE-MAR | Marburg | Germany | 2007 |
| Robert Koch Institute | DE-BER | Berlin | Germany | 2013 |
| Azienda Ospedaliera Ospedale Luigi Sacco | ITA-MIL | Lombardy, Milano | Italy | >2007 |
| Istituto Nazionale Malattie Infettive | ITA-INMI | Rome | Italy | 2012 |
| Netherlands National Institute for Public Health and the Environment (RIVM) | NL-RIVM | Bilthoven | Netherlands | 2010 |
| Swedish Institute for Communicable Disease Control | SW-SMI | Solna | Sweden | 2001 |
| High Containment Laboratory DDPS (SiLab) | CH-SILAB | Spiez | Switzerland | 2011 |
| University of Geneva (P4D) | CH-HUG | Geneva | Switzerland | 2007 |
| Defence Science and Technology Laboratory | UK-DSTL | Porton Down, Wiltshire | UK | 2005 |
| Centre for Emergency Preparedness and Response, Health Protection Agency (HPA) | UK-HPA-SPRU | Porton Down, Wiltshire | UK | <1987 |
| Health Protection Agency's Centre for Infections | UK-HPA-CIS | Colindale | UK | <1987 |
| National Institute for Biological Standards and Control (NIBSC) | UK-NIBSC | Potters Bar, Hertfordshire | UK | 1987 |
| Veterinary Laboratories Agency | UK-VLA | Addlestone, Surrey | UK | <2003 |
| Institute for Animal Health | UK-IAH | Pirbright | UK | 2006 |
| Merial Animal Health Ltd | UK-MER | Pirbright | UK | 2007 |
| National Institute for Medical Research | UK-NIMR | London | UK | 2006 |
| Schering-Plough Animal Health | UK-SP | Harefield | UK | 2007 |
| Centre International de Recherches Médicales de Franceville | GA-CIRMF | Franceville | Gabon | 1998 |
| National Institute for Communicable Diseases | ZA-NICD | Johannesburg | South Africa | 1980 |

*Geographical coordinate of each facility can be requested directly to the authors.